\documentstyle[pmart,twoside,fleqn,epsf]{article} % Do not change this line!

\begin{document}

%
%  The title should be in capital letters!
%
\title{Quantum Phase Transition \\
       in the One-Dimensional XZ Model }

\author{Wojciech Brzezicki $^{a}$ and Andrzej M. Ole\'s $^{a,b}$}

\address{$^{a}$ Marian Smoluchowski Institute of Physics,
                Jagellonian University  \\
                Reymonta 4, PL--30059 Krak\'ow, Poland \\
         $^{b}$ Max-Planck-Institut FKF, Heisenbergstrasse 1,
                D--70569 Stuttgart, Germany }

\date{April 29, 2008}

\maketitle

\pacs{75.10.Jm, 03.67.-a, 05.70.Fh, 63.70.Tg}

\begin{abstract}
We introduce a one-dimensional (1D) XZ model with alternating
$\sigma_i^z\sigma_{i+1}^z$ and $\sigma_i^x\sigma_{i+1}^x$
interactions on even/odd bonds, interpolating between the Ising
model and the quantum compass model. We present two ways of its
exact solution by: ($i$) mapping to the quantum Ising models, and
($ii$) using fermions with spin $1/2$. In certain cases the
nearest neighbor pseudospin correlations change discontinuously at
the quantum phase transition, where one finds highly degenerate
ground state of the 1D compass model.
\end{abstract}

{\em Introduction---} Models of magnetism with exotic interactions
are motivated by rather complex orbital superexchange in Mott
insulators. In certain case the degeneracy of $3d$ orbitals is
only partly lifted and the remaining orbital degrees of freedom
are frequently described as $1/2$ spins. They also arise from
spin-orbital superexchange, with rich dynamics leading to enhanced
quantum fluctuations near quantum phase transitions \cite{Fei},
and to entangled spin-orbital ground states \cite{Ole}. The
orbital interactions have much lower symmetry than the SU(2) of
spin interactions and their form depends on the orientation of the
bond in real space \cite{vdB}, so they may lead to orbital liquid
in three dimensions \cite{Kha}. A generic and simplest model of
this type is so-called compass model introduced in \cite{Kho} when
the coupling along a given bond is Ising-like, but {\em different}
spin components are active along particular bonds, for instance
$J_x\sigma_i^x\sigma_j^x$ and $J_z\sigma_i^z\sigma_j^z$ along $a$
and $b$ axis in the two-dimensional (2D) compass model
\cite{Mil05}. This situation is quite different from classical
Ising-like models with periodically distributed frustration
\cite{Lon80}.

Recently we discussed the properties of the one-dimensional (1D)
model that interpolates between Ising and compass model \cite{My}.
The solution based on specific choice of interactions indicated
divergences in correlation functions while approaching the
transition point. This suggested first order phase transition but
it was not clear if the effect is generic or it is just an
artefact of singular parametrization of interactions. To answer
this question we introduce more general solution which can be
applied to any parametrization and then we make second insight
into the original problem.

{\em XZ model in one dimension---} The model is described by a
generalized XZ Hamiltonian with different energies for even and
odd bonds, where the number of sites is $2N$ and for simplicity we
suppose that $N$ is even
\begin{equation}
{\cal H}=\sum_{i=1}^{N}\left\{
J_{1}\sigma_{2i-1}^z\sigma_{2i}^z+J_{2}\sigma_{2i-1}^x\sigma_{2i}^x
+L_{1}\sigma_{2i}^z\sigma_{2i+1}^z+L_{2}\sigma_{2i}^x\sigma_{2i+1}^x
\right\}\ .
\label{ham}
\end{equation}
This Hamiltonian turns into the one discussed in Ref. \cite{My} if
we fix the energy constants as follows:
$J_{1}(\alpha)=\frac{1}{2}(|1-\alpha|+1-\alpha)$,
$J_{2}(\alpha)=-J_{1}(\alpha)+1$,
$L_{1}(\alpha)=-J_{1}(-\alpha+2)+1$,
$L_{2}(\alpha)=J_{1}(-\alpha+2)$, where $0\leq\alpha\leq2$. For
$0\leq\alpha\leq1$ it gives
\begin{equation}
{\cal H}(\alpha)\equiv \sum_{i=1}^{N} \left\{ (1-\alpha)
\sigma_{2i-1}^z\sigma_{2i}^z +\alpha\sigma_{2i-1}^x\sigma_{2i}^x
+\sigma_{2i}^z\sigma_{2i+1}^z \right\}\ ,
\label{Ha1}
\end{equation}
and for $1<\alpha\leq2$ we transform $\sigma_{i}^x\leftrightarrow
\sigma_{i}^z$, $\{2i-1,2i\}\leftrightarrow\{2i,2i+1\}$ for all $i$
and $\alpha\rightarrow(2-\alpha)$.

\begin{figure}[t!]
\epsfysize=4.4cm \centerline{\epsfbox{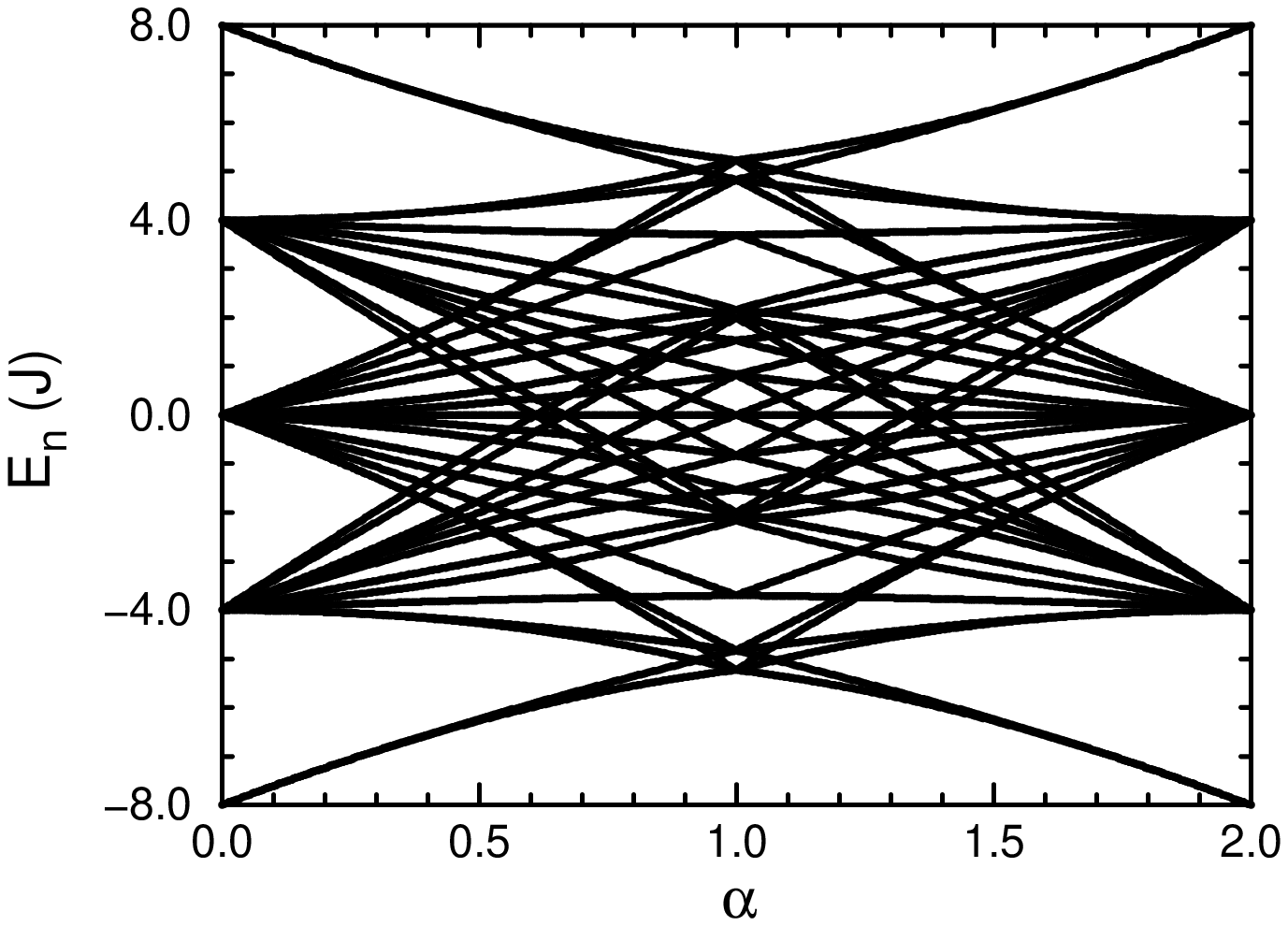} \hskip .6cm
\epsfysize=4.4cm \epsfbox{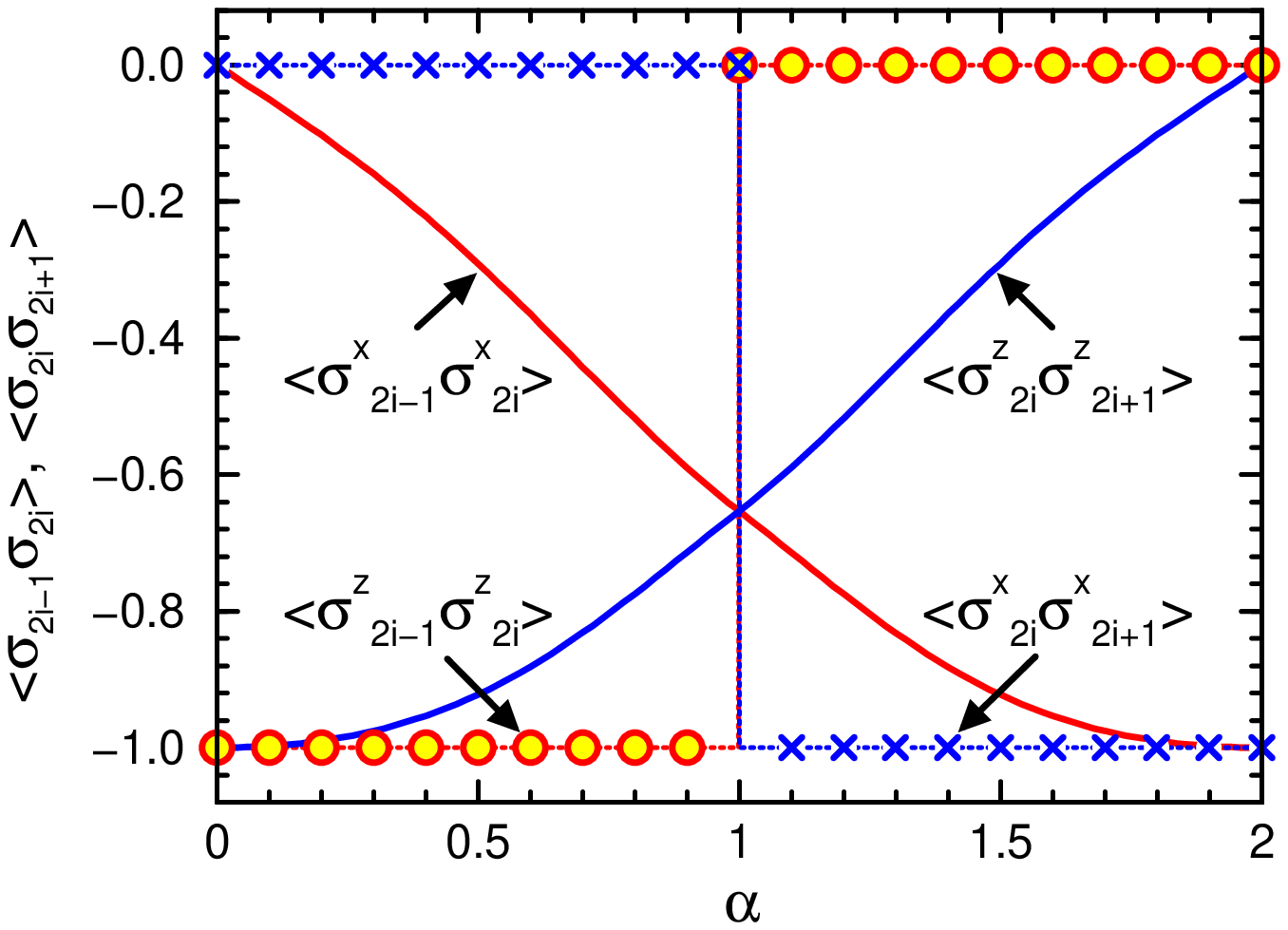}} \leftline{\hskip 3.2cm
(a)\hskip 6.1cm (b)} \caption{ (a) Eigenenergies $E_n$ of the
$XX$--$ZZ$ model (\ref{Ha1}) for $2N=8$ sites and for increasing
$\alpha$. Level crossing at $\alpha=1$ marks the quantum critical
point of the 1D compass model. (b) Intersite pseudospin
correlations on odd $\{2i-1,2i\}$ and even $\{2i,2i+1\}$ bonds in
the $XX$--$ZZ$ model (\ref{Ha1}) for increasing $\alpha$. A
transition between two types of pseudo-order, with
$\langle\sigma_{2i-1}^z\sigma_{2i}^z\rangle=-1$ for $\alpha<1$ and
$\langle\sigma_{2i}^x\sigma_{2i+1}^x\rangle=-1$ for $\alpha>1$,
occurs at the quantum critical point $\alpha=1$, where only
$\langle\sigma_{2i-1}^x\sigma_{2i}^x\rangle=
 \langle\sigma_{2i}^z\sigma_{2i+1}^z\rangle=-\frac{2}{\pi}$
are finite. These results are reproduced from Ref. \cite{My}. }
\label{fig:com}
\end{figure}

{\em First solution---} To solve the model given by Eq.
(\ref{Ha1}) we choose eigenbasis of $\sigma_i^z$ operators
consisting of vectors
$\left|s_{1},s_{2},s_{3},\ldots,s_{2N}\right\rangle$ with
$s_i=\pm1$ for all $i$. Every state like that can be denoted
equivalently as
\begin{equation}
\left|t_{1},t_{2},\ldots,t_{N}\right\rangle _{r_{1}r_{2}\ldots
r_{N}}
\equiv\left|t_{1},t_{1}r_{1},t_{2},t_{2}r_{2},\ldots,t_{N},t_{N}r_{N}
\right\rangle\ , \label{sub}
\end{equation}
where $t_{i}\equiv s_{2i-1}$ and $r_{i}\equiv s_{2i-1}s_{2i}$ for
$i=1,2,\dots,N$. This will let us exploit the fact that the
hamiltonian (\ref{Ha1}) flips only odd pairs of spins. For states
like (\ref{sub}), we define new spin operators $\tau_{i}^{'z}$ and
$\tau_{i}^x$ which act only on $t_1,t_2,\dots,t_N$ quantum numbers
\begin{eqnarray}
\tau_{1}^x |t_{1},t_{2},\ldots,t_{N}\rangle _{r_{1}r_{2}\cdots
r_{N}}& = & |-t_{1},t_{2},\ldots,t_{N}\rangle _{r_{1}r_{2}\cdots
r_{N}}\ ,
\nonumber \\
\tau_{1}^{'z} |t_{1},t_{2},\ldots,t_{N}\rangle _{r_{1}r_{2}\cdots
r_{N}}& = & t_{1} |t_{1},t_{2},\ldots,t_{N}\rangle
_{r_{1}r_{2}\cdots r_{N}}\ .
\end{eqnarray}
Next we transform each $\tau_{i}^{'z}$ as follows
$\tau_{i}^{'z}\equiv r_{1}r_{2}\cdots r_{i-1}\tau_{i}^z$, to get
effective forms of the Hamiltonian in subspaces spanned by vectors
(\ref{sub}) with fixed $r_i$'s
\begin{equation}
{\cal H}_{r_{1}r_{2}\cdots r_{N}}(\alpha)=\sum_{i=1}^{N}\left\{
\tau_{i}^z\tau_{i+1}^z+\alpha \tau_{i}^x\right\}+C_{\vec
r}(\alpha)\ ,
\label{hamef}
\end{equation}
where $\tau_{N+1}^z=r_{1}r_{2}\dots r_{N}\tau_{1}^z$ and $C_{\vec
r}(\alpha)=(1-\alpha)\sum_{i=1}^Nr_i$. Now we have to solve two
types of quantum Ising model (QIM); either with periodic or
antiperiodic boundary condition. The solution is well known and
was described in detail in Ref. \cite{My}. First step is to
introduce Jordan-Wigner (JW) transformation,
\begin{eqnarray}
%\begin{array}{rcl}
\tau_{j}^z & = & (c_{j}^{}+c_{j}^{\dagger})
{\prod_{i<j}}(1-2c_{i}^{\dagger}c_{i}),\nonumber \\
\tau_{j}^x & = & (1-2c_{j}^{\dagger}c_{j}^{}).
%\end{array}
\end{eqnarray}
The boundary condition for fermion operators $\{c_i\}$ differs for
even and odd number of quasiparticles in the chain. Fortunately,
the Hamiltonian conserves the parity of fermions, this lets us
write ${\cal H}_{\vec r}=\frac{1}{2}(1+P){\cal H}_{\vec
r}^++\frac{1}{2}(1-P){\cal H}_{\vec r}^-$, where
\begin{equation}
{\cal H}_{\vec r}^{\pm}=\sum_{i=1}^N\{(c_{i}^{\dagger}-c_{i})
(c_{i+1}^{\dagger}-c_{i+1})-2\alpha c_{i}^{\dagger}c_{i}\}+C_{\vec
r}(\alpha)\ , \label{qimJW}
\end{equation}
and where $\frac{1}{2}(1\pm P)$ are projections on subspaces with
even (+) and odd ($-$) number of quasiparticles. Here we adopt
notation $\vec r$ for the subspace labels $r_1r_2\dots r_N$. The
boundary conditions are $c_{N+1}=\mp c_1\prod_{i=1}^Nr_i$ for
${\cal H}_{\vec r}^{\pm}$ respectively. After the Fourier
transformation, $c_{j} =
\frac{1}{\sqrt{N}}{\sum_{k}}^{\pm}e^{ijk}c_{k}$ we find
(\ref{qimJW}) in a block diagonal form
\begin{equation}
{\cal H}_{\vec r}^{\pm}={\sum _k}^{\vec r \pm} \{c^\dagger _k
c_k(2\cos k -2\alpha)+ (c^\dagger _kc^\dagger
_{-k}e^{ik}+h.c.)\}+C_{\vec r}(\alpha)+N\alpha\ ,
\end{equation}
where quasimomenta $k$ take ``integer'' values $k=0,\pm
\frac{2\pi}{N},\pm 2 \frac{2\pi}{N},\dots,\pi$ for a periodic
boundary condition and in the antiperiodic case they are
``half-integer'' $k=0,\pm \frac{2\pi}{N},\pm 2
\frac{2\pi}{N},\dots,\pi$. Diagonalization is completed by a
Bogoliubov transformation
$\gamma_k^{\dagger}=u_kc_k^{\dagger}+v_{k}c_{-k}$ preformed for
all $k>0$ and $k\not=\pi$ where  $(u_k,v_k)$ are eigenmodes of the
Bogoliubov-de Gennes equation $[{\cal H}_{\vec
r}^{\pm},\gamma_k^{\dagger}]=E_k\gamma_k^{\dagger}$. In this way
we get the full Hamiltonian's spectrum in every subspace $\vec r$.
For instance, in case when $\prod_{i=1}^Nr_i=1$ we obtain a
Hamiltonian for even number of quasiparticles $\gamma_k$
\begin{equation}
{\cal H}_{\vec r}^{+}(\alpha)={\sum _k}^{\vec r +}E_k
\left(\gamma_k^{\dagger}\gamma_k - \frac{1}{2}\right)
+C_{\vec
r}(\alpha),
\end{equation}
where $E_k=2\{1+\alpha^2-2\alpha\cos{k}\}^{1/2}$ is the
quasiparticle energy. The ground state is Bogoliubov vacuum in a
subspace where all $r_i=-1$, apart from points $\alpha=0,1,2$ it
has no degeneracy. Cases of $\alpha=1,2$ are trivial. For
$\alpha=1$ we have $C_{\vec r}(\alpha)=0$ for all $\vec r$ which
means that there is a ground state in every subspace where
$\prod_{i=1}^N r_i=1$. This results in $2^{N-1}$-fold degeneracy
for the 1D compass model [Fig. \ref{fig:com}(a)]. In the limit
$N\to\infty$ the lowest energies of periodic and antiperiodic QIM
get equal at $\alpha=1$. For $0<\alpha\leq 1$ they are already
two-- and threefold degenerate, so when $\alpha=1$ the total
degeneracy is $5\times2^{N}$, and the spin gap vanishes \cite{My}.

%%%%%%%%%%%%%%%%%%%%%%%%%%%%%%%%%%%%%%%%
%%%%%%%%%%Figure%%%%%%%%%%%%%%%%%%%%%%%%
%%%%%%%%%%%%%%%%%%%%%%%%%%%%%%%%%%%%%%%%

%%%%%%%%%%%%%%%%%%%%%%%%%%%%%%%%%%%%%%%%
%%%%%%%%%%Second way%%%%%%%%%%%%%%%%%%%%
%%%%%%%%%%%%%%%%%%%%%%%%%%%%%%%%%%%%%%%%

{\em Second solution---} The most direct way of dealing with Eq.
(\ref{ham}) is to leave the interactions $J_{1,2}$ and $L_{1,2}$
undefined and to start with the JW transformation
\begin{eqnarray}
%\begin{array}{rcl}
\sigma_{j}^z & = & (c_{j}+c_{j}^{\dagger})
{\prod_{i<j}}(1-2c_{i}^{\dagger}c_{i}),\nonumber\\
\sigma_{j}^x & = & \frac{1}{i}(c_{j}-c_{j}^{\dagger})
{\prod_{i<j}}(1-2c_{i}^{\dagger}c_{i}),
%\end{array}
\end{eqnarray}
which transforms spins into fermion operators $c_{j}$. Here the
crucial step is to introduce new quantum number for fermions with
two possible values $n$ and $p$. This can be regarded as
quasiparticles' spin or as splitting the chain into bi-atomic
elementary cells. We define $c_{i}^{n}\equiv c_{2i-1}$ and
$c_{i}^{p}\equiv c_{2i}$. Because of the boundary conditions and
the fact that $\cal H$ preserves the parity of fermions, we obtain
\begin{equation}
{\cal
H}^{\pm}=\sum_{i=1}^{N}\left\{J^{+}c_{i}^{n\dagger}c_{i}^{p}+L^{+}
c_{i}^{p\dagger}c_{i+1}^{n}+J^{-}c_{i}^{n\dagger}c_{i}^{p\dagger}
+L^{-}c_{i}^{p\dagger}c_{i+1}^{n\dagger}+h.c.\right\}\ ,
\label{HAMnp}
\end{equation}
where ${\cal H}^+$ (${\cal H}^-$) is defined in subspace with even
(odd) number of fermions with $c_{N+1}^n=-c_1^n$
($c_{N+1}^n=c_1^n$). Here we introduced new notation for $J$'s;
$J^{\pm}=J_{1}\pm J_{2}$ and analogically for $L$'s. Now, we
proceed with Fourier transform $c_{j}^{n,p}=
\frac{1}{\sqrt{N}}{\sum_{k}}^{\pm}e^{ijk}c_{k}^{n,p}$ for $n$ and
$p$ fermions separately, compatible to fermions' parity. For
periodic boundary conditions $k$'s take ``integer'' values  and in
the antiperiodic case they are ``half-integer''. Finally we find
the problem block diagonal in a form
\begin{equation}
{\cal H}^{+}={\sum_k}^{+}\left\{c_{k}^{n\dagger}c_{-k}^{p\dagger}
(J^{-}-L^{-}e^{-ik}))
+c_{k}^{n\dagger}c_{k}^{p}(J^{+}+L^{+}e^{-ik})+h.c.\right\}\ ,
\end{equation}
and for ${\cal H}^{-}$ similarly but with $k=0,\pi$ possible.
Diagonalization is completed by a four-dimensional Bogoliubov
transformation. We search for invariant subspace of a linear
operator $[{\cal H}^{+},.]$ in 8-dimensional space spanned by
$c_{\pm k}^{n(\dagger)}$ and $c_{\pm k}^{p(\dagger)}$. The result
suggests the form of transformation as
\begin{equation}
\begin{array}{ccc}
\left(\begin{array}{c}
\gamma_{k}^{n\dagger}\\
\gamma_{-k}^{n}\\
\gamma_{k}^{p\dagger}\\
\gamma_{-k}^{p}\end{array}\right)={\hat\beta}_{k}\left(\begin{array}{c}
c_{k}^{n\dagger}\\
c_{-k}^{n}\\
c_{k}^{p\dagger}\\
c_{-k}^{p}\end{array}\right) & (0\le k<\pi),\hskip .3cm &
\left(\begin{array}{c}
\gamma_{\pi}^{n\dagger}\\
\gamma_{\pi}^{p\dagger}\end{array}\right)=
{\hat\beta}_{\pi}\left(\begin{array}{c}
c_{\pi}^{n\dagger}\\
c_{\pi}^{p\dagger}\end{array}\right)\end{array},
\end{equation}
where ${\hat\beta}_{k}$ and ${\hat\beta}_{\pi}$ are orthogonal
matrices. This assures that $\gamma_k$'s are fermionic. The rows
of  ${\hat\beta}$ are eigenvectors of Bogoliubov-de Gennes
equation $[{\cal H},\gamma_{k}^{n\dagger}]
=E_{k}\gamma_{k}^{n\dagger}$ and eigenvalues $E_k$ are energies of
sytem's elementary excitations. Finally, we find ${\cal H}^+$ in a
diagonal form
\begin{equation}
{\cal H}^{+}= {\sum_{k}}^{+} \left\{E_{k}^{n}
\left(\gamma_{k}^{n\dagger}\gamma_{k}^{n}-\frac{1}{2}\right)+
E_{k}^{p}
\left(\gamma_{k}^{p\dagger}\gamma_{k}^{p}-\frac{1}{2}\right)
\right\}\ , \label{npdiag}
\end{equation}
where $E_{k}^{n}=2\{J_{1}^{2}+L_{2}^{2}+2 J_{1}L_{2}\cos
k\}^{1/2}$, $E_{k}^{p}=2\{J_{2}^{2}+L_{1}^{2}+2 J_{2}L_{1}\cos
k\}^{1/2}$. In a similar way we get the result for ${\cal H}^-$.
Luckily, the parity of particles $\gamma_k$ is the same as parity
of original JW fermions, thus only states with even (odd) number
of quasiparticles $\gamma_k$ belong to the spectrum of ${\cal
H}^+$ (${\cal H}^-$). The ground state energy $E_0$ obtained from
(\ref{npdiag}) is
$E_0=-\frac{1}{2}{\sum_k}^{+}[E_{k}^{n}+E_{k}^{p}]$. Putting
$J_{1,2}=J_{1,2}(\alpha)$ and $L_{1,2}=L_{1,2}(\alpha)$ we find
the same energy spectrum as described in Ref. \cite{My} [see Fig.
\ref{fig:com}(a)]. For $\alpha\leq 1$ one finds
$E_{k}^{n}=2(1-\alpha)$ and
$E_{k}^{p}=\{1+\alpha^2-2\alpha\cos{k}\}^{1/2}$, which means that
the occupation numbers $\gamma_{k}^{n\dagger}\gamma_{k}^{n}$ play
the role of the subspace indexes $r_i$ from the previous solution
while $\gamma_{k}^{p\dagger}\gamma_{k}^{p}$ describe excitations
within a given subspace.

Pseudospin correlation functions can be derived from $E_0$ as
derivatives with respect to $J_{1,2}$ and $L_{1,2}$, respectively.
The main result is that
$\langle\sigma_{2i-1}^z\sigma_{2i}^z\rangle$ and
$\langle\sigma_{2i}^x\sigma_{2i+1}^x\rangle$ remain constant in
intervals $[0,1)$ and $(1,2]$, but with discontinuities at
$\alpha=1$ [see Fig. \ref{fig:com}(b)]. The origin of these
singularities is a cusp of $E_k^n(J_1,L_2)$ surface at
$(J_1,L_2)=(0,0)$. The trajectory $(J_1,L_2)(\alpha)$ passes
through this point at $\alpha=1$. This means that no first order
phase transition occurs in the general model of Eq. (\ref{ham}),
unless curves $(J_1,L_2)(\alpha)$ or $(J_2,L_1)(\alpha)$ pass
through $(0,0)$. Passing should be interpreted literally as
passing, not reversing at $(0,0)$. For example, the curve
$(x,y)=((\alpha-1)^2,(\alpha-1)^3)$ reaches the point $(0,0)$ at
$\alpha=1$, but nevertheless one of the pseudospin correlations
remains continuous.

{\em Summary---} We have presented an exact solution which
demonstrated that a hidden order with constant pseudospin
correlations exists in the 1D XZ model. The second method used for
solving this problem provides more insight into the nature of the
quantum phase transition as discussed in Ref. \cite{My}, while the
second one is more flexible and may be generalized, for instance,
to the ladder geometry.

\section*{Acknowledgments}

A.M. Ole\'s acknowledges support by the Foundation for Polish
Science (FNP). This work was supported by the Polish Ministry of
Science and Education under Project No.~N202 068 32/1481.

\end{document}